\documentclass[twocolumn,showpacs,preprintnumbers,amsmath,amssymb,showpacs,showcolor]{revtex4}

\usepackage{graphicx}
\usepackage{dcolumn}
\usepackage{bm}
\usepackage{amssymb}
\usepackage{color}

 \def\tskip{\setlength{\tskip}{5pt}}
\def\colwidth{\setlength{\colwidth}{3.5in}}

\def\prd{Phys. Rev. D}
\def\pr{Phys. Rep.~}
\def\prl{Phys. Rev. Lett.~}
\def\plb{Phys. Lett. B}
\def\jcap{JCAP~}
\def\apj{Astrophys. J.~}

\def\mnras{Mon. Not. Roy. Astron. Soc.~}
\def\apjs{Astrophys. J. Suppl. Ser.~}

\def\cqg{Class. Quant. Grav. ~}
\def\lnp{Lect. Notes Phys.~}

\def\ijmpd{Int. J. Mod. Phys. D}

\newcommand{\lsim}{\mathrel{\hbox{\rlap{\lower.55ex\hbox{$\sim$}} \kern-.3em \raise.4ex \hbox{$<$}}}}
\newcommand{\gsim}{\mathrel{\hbox{\rlap{\lower.55ex\hbox{$\sim$}} \kern-.3em \raise.4ex \hbox{$>$}}}}
\newcommand{\beq}{\begin{equation}}
\newcommand{\eeq}{\end{equation}}
\newcommand{\be}{\begin{equation}}
\newcommand{\ee}{\end{equation}}
\newcommand{\bes}{\begin{equation*}}
\newcommand{\ees}{\end{equation*}}
\newcommand{\beqa}{\begin{eqnarray}}
\newcommand{\eeqa}{\end{eqnarray}}
\newcommand{\bea}{\begin{eqnarray}}
\newcommand{\ena}{\end{eqnarray}}

\newcommand{\mpl}{m_\mathrm{Pl}}

\def\mpl{m_{\mathrm{Pl}}}

\begin{document}

\title{Constraints of Relic Gravitational Waves by Pulsar Timing Array: Forecasts for the FAST and SKA Projects}

\author{Wen Zhao$^{1,2}$, Yang Zhang$^{1,2}$, Xiao-Peng You$^{3}$ and Zong-Hong Zhu$^{4}$}
\affiliation{$^{1}$Department of Astronomy, University of Science
and Technology of China, Hefei, 230026, China \\ $^{2}$Key
Laboratory for Researches in Galaxies and Cosmology, University of
Science and Technology of China, Hefei, 230026, China \\
$^{3}$School of Physical Science and Techology, Southwest
University, Chongqing, 400715, China \\ $^{4}$Department of
Astronomy, Beijing Normal University, Beijing 100875, China}


\begin{abstract}

Measurement of the pulsar timing residuals provides a direct way
to detect relic gravitational waves at the frequency $f\sim 1/{\rm
yr}$. In this paper, we investigate the constraints on the
inflationary parameters, the tensor-to-scalar ratio $r$ and the
tensor spectral index $n_t$, by the current and future Pulsar
Timing Arrays (PTAs). We find that Five-hundred-meter Aperture
Spherical radio Telescope (FAST) in China and the planned Square
Kilometer Array (SKA) projects have the fairly strong abilities to
test the phantom-like inflationary models. If $r=0.1$, FAST could
give the constraint on the spectral index $n_t<0.56$, and SKA
gives $n_t<0.32$. While an observation with the total time
$T=20$yr, the pulsar noise level $\sigma_w=30$ns and the monitored
pulsar number $n=200$, could even constrain $n_t<0.07$. These are
much tighter than those inferred from the current results of
Parkers Pulsar Timing Array (PPTA), European Pulsar Timing Array
(EPTA) and North American Nanohertz Observatory for Gravitational
waves (NANOGrav). Especially, by studying the effects of various
observational factors on the sensitivities of PTAs, we found that
compared with $\sigma_w$ and $n$, the total observation time $T$
has the most significant effect.

\end{abstract}



\maketitle


\section{Introduction \label{section1}}

In a whole range of scenarios of the early Universe, including the
well-studied inflationary models, a stochastic background of relic
(primordial) gravitational waves (RGWs) was produced due to the
superadiabatic amplification of zero point quantum fluctuations of
the gravitational field
\cite{grishchuk1974,starobinsky1979,mukhanov1992}. Their detection
maybe provide the unique way to study the birth of the Universe,
the expansion history of Universe before the recombination stage,
and test the applicability of general relativity and quantum
mechanics in the extremely high-energy scale \cite{grishchuk2007}.

Since RGWs have a wide range spreading spectra, from $10^{-18}$Hz
to $10^{10}$Hz, one can detect or constrain them at different
frequencies. The temperature and polarization anisotropies of
Cosmic Microwave Background (CMB) radiation provide the way to
constrain RGWs at very low frequencies, $f<10^{-15}$Hz. Nowadays,
combining with other cosmological observations, the nine-year WMAP
data place the constraint on the tensor-to-scalar ratio $r<0.13$
\cite{wmap9}. While the new Planck data give the the tightest
constraint $r<0.11$ \cite{planck2013}, which is equivalent to the
constraint of the amplitude of RGWs at lowest frequency $f\sim
10^{-17}$Hz. In the near future, this bound will be greatly
improved by the forthcoming polarization observations of Planck
satellite, several ground-based and balloon-borne experiments
(BICEP, QUIET, POLARBEAR, QUIJOTE, ACTPOL, SPTPOL, QUBIC, EBEX,
PIPER, SPIDER et al.), and the planned fourth-generation CMB
missions (CMBPol, LiteBird, COrE. et al.).

Among all the direct observations, LIGO S5 has experimentally
obtained so far the most stringent bound $\Omega_{\rm gw}\le
6.9\times10^{-6}$ around $f\sim 100$Hz \cite{ligo2009,zhang2010}.
It is expected that AdvLIGO, AdvVIRGO, KAGRA,  ET and eLISA will
also deeply improve it in the near future. In particular, the
Planned BBO, DECIGO and ASTROD projects may directly detect the
signal of RGWs in the far future.  In addition, there are two
bounds on the integration $\int \Omega_{\rm gw}(f)d\ln f
\lsim1.5\times 10^{-5}$, obtained by the Big Bang nucleosynthesis
(BBN) observation \cite{bbn1999} and the CMB observation
\cite{cmb2006}.

By analyzing of pulsar pulse time-of-arrival (TOA) data, people
find the millisecond pulsars are very stable clocks. The
measurement of their timing residuals provides a direct way to
detect GW background in the frequency range $f\in
(10^{-9},~10^{-7})$Hz \cite{detweiler1979,hellings,pulsarreview}.
In addition to the GWs generated by the coalescence of massive
black hole binary systems \cite{blackhole} and cosmic strings
\cite{cosmicstring}, RGWs are another kind of most important GW
sources in this frequency range. Recently, PPTA, EPTA and NANOGrav
teams have reported their observational results on the stochastic
background of GWs. In \cite{zhaopulsar}, by considering these
results, we have detailedly investigated the constraints on the
Hubble parameter during inflation in the most general scenario for
the early Universe. In this paper, we shall extend them to
constrain the tensor-to-scalar ratio $r$ and the tensor spectral
index $n_t$.

In addition, as the main goal of this paper, we will discuss the
potential constraint (or detection) of RGWs by the future PTA
observations. In our discussion, FAST and SKA will be treated as
two typical projects, and mainly focused on in the studies. The
dependence of the RGWs constraints on the total observation time
$T$, the number of monitored pulsars $n$ and the magnitude of the
pulsar timing noise $\sigma_w$ will be discussed.

This paper is constructed as follows, In Sec. 2, we briefly review
the model to describe the RGWs, and relate the energy density of
GWs $\Omega_{\rm gw}$, the tensor-to-scalar ratio $r$ and the
tensor spectral index $n_t$ to the characteristic strain spectrum
$h_c(f)$, which is widely used in the PTA analysis. In Sec. 3, we
describe the sensitivities of the current and future experiments,
and discuss the dependence on various observational parameters.
Sec. 4 summaries the main results of this paper.


\section{Relic gravitational waves \label{section2}}

Incorporating the perturbation to the spatially flat
Friedmann-Robertson-Walker spacetime, the metric is
 \beq\label{metric}
 ds^2=a^{2}(\eta) \left[d\eta^2-(\delta_{ij}+h_{ij}dx^idx^j)\right],
 \eeq
where $a$ is the scale factor of the universe, and $\eta$ is the
conformal time, which relates to the cosmic time by $a d\eta=dt$.
The perturbation of spacetime $h_{ij}$ is a $3\times 3$ symmetric
matrix. The gravitational-wave field is the tensorial portion of
$h_{ij}$, which is transverse-traceless $\partial_ih^{ij}=0$,
$\delta^{ij}h_{ij}=0$.

{RGWs satisfy the linearized evolution equation
\cite{grishchuk1974}:
 \beq\label{evolution}
 \partial_{\mu}(\sqrt{-g}\partial^{\mu}h_{ij})=-16\pi G \pi_{ij}.
 \eeq
The anisotropic portion $\pi_{ij}$ is the source term, which can
be given by the relativistic free-streaming gas
\cite{weinberg2003}. However, it has been deeply discussed that
the relativistic free-streaming gas, such as the decoupled
neutrino, can only affect the RGWs at the frequency range
$f\in(10^{-16},~10^{-10})$Hz, which could be detected by the
future CMB observations \cite{zzx2009}. So, it cannot obviously
influence the RGWs at the frequency $f\in (10^{-9},~10^{-7})$Hz}.
For this reason, in this paper we shall ignore the contribution of
the external sources. So the evolution of RGWs only depends on the
scale factor and its time derivative.

It is convenient to Fourier transform the equation as follows:
 \be\label{fourier}
 h_{ij}(\eta,\vec{x})=\int \frac{d^3 \vec{k}}{(2\pi)^{{3}/{2}}}\sum_{s=+,\times}
 \left[
 h_k(\eta) \epsilon^{(s)}_{ij}c^{(s)}_{\vec{k}}e^{i\vec{k}\cdot\vec{x}}+c.c.
 \right],
 \ee
where $c.c.$ stands for the complex conjugate term. The
polarization tensors are symmetry, transverse-traceless
$k^i\epsilon_{ij}^{(s)}(\vec{k})=0$,
$\delta^{ij}\epsilon_{ij}^{(s)}(\vec{k})=0$, and satisfy the
conditions
$\epsilon^{(s)ij}(\vec{k})\epsilon_{ij}^{(s')}(\vec{k})=2\delta_{ss'}$
and $\epsilon_{ij}^{(s)}(-\vec{k})=\epsilon_{ij}^{(s)}(\vec{k})$.
Since the RGWs we will consider are isotropy, and each
polarization state is the same, we have denoted
$h_{\vec{k}}^{(s)}(\eta)$ by $h_k(\eta)$, where $k=|\vec{k}|$ is
the wavenumber of the GWs, which relates to the frequency by
$k\equiv 2\pi f$. (The present scale factor is set $a_0=1$). So
Eq.(\ref{evolution}) can be rewritten as
 \beq\label{h-evolution}
{h_k}''+2\frac{{a'}}{a}{h_k}'+k^2h_k=0,
 \eeq
where the {\it prime} indicates a conformal time derivative
$d/d\eta$. For a given wavenumber $k$ and a given time $\eta$, we
can define the transfer function $t_f$ as
 \be\label{tf-define}
 t_f(\eta,k)\equiv h_k(\eta)/h_k(\eta_i),
 \ee
where $\eta_i$ is the initial time. This transfer function can be
obtained by solving the evolution equation (\ref{h-evolution}).

The strength of the GWs is characterized by the GW energy
spectrum, $\Omega_{\rm gw} \equiv \rho_{\rm gw}/\rho_0$, where
$\rho_{\rm gw}=\frac{1}{32\pi
G}\langle\dot{h}_{ij}\dot{h}^{ij}\rangle$, the critical density is
$\rho_0=\frac{3H_0^2}{8\pi G}$, and $H_0$ is the current Hubble
constant. Using Equations in (\ref{fourier}) and
(\ref{tf-define}), the energy density of GWs can be written as
\cite{page2006}
 \be
 \rho_{\rm gw}=\int \frac{d k}{k} \frac{P_t(k)\dot{t}^2_f(\eta_0,k)}{32\pi G},
 \ee
where $P_t(k)\equiv \frac{2k^3}{\pi^2}|h_k(\eta_i)|^2$ is the
so-called primordial power spectrum of RGWs. Thus, we derive that
the current energy density of RGWs,
 \be\label{omega-gw}
 \Omega_{\rm gw} \equiv \int \Omega_{\rm gw} (k)d\ln{k} ,
{~\rm and~}
 \Omega_{\rm gw} (k) = \frac{P_t(k)}{12
 H_0^2}\dot{t}_f^2(\eta_0,k),
 \ee
where the  {\it dot} indicates a cosmic time derivative $d/dt$.

Now, let us discuss the terms $P_t(k)$ and $t_f(\eta_0,k)$
separately. The primordial power spectrum of RGWs is usually
assumed to be power-law as follows:
 \be\label{pt1}
 P_t(k) = A_t(k_*)\left(\frac{k}{k_*}\right)^{n_t}.
 \ee
This is a generic prediction of a wide range of scenarios of the
early Universe, including the inflation models. $A_t(k_*)=\frac{16
H^2_*}{\pi\mpl^2}$ directly relates to the value of Hubble
parameter $H$ at time when wavelengths corresponding to the
wavenumber $k_*$ crossed the horizon
\cite{grishchuk1974,mukhanov1992,peiris2003,zhaopulsar}. In
observations, we always define the tensor-to-scalar ratio $r$, and
write the amplitude of RGWs as $A_t(k_*)=A_s r$, where $A_s$ is
the amplitude of primordial density perturbation at $k=k_*$. $n_t$
is the spectral index of RGWs, which relates to the effective
equation-of-state $w$ of the cosmic ``matter" in the inflationary
stage by the relation,
 \be
 n_t=\frac{4}{1+3w}+2.
 \ee
If the inflation is an exact de Sitter expansion stage with
$w=-1$, we have the scale-invariant spectrum with $n_t=0$. For the
canonical scalar-field inflationary models, we have $w>-1$, which
predicts the red spectrum of RGWs with $n_t<0$
\cite{mukhanov1992}. However, for the phantom inflationary models
\cite{phantom}, one has $w<-1$ and $n_t>0$. So the determination
of $n_t$ can distinguish different kinds of inflationary
scenarios.

Now, let us turn to the transfer function $t_f$, defined in
(\ref{tf-define}), which describes the evolution of GWs in the
expanding Universe. From Eq.(\ref{h-evolution}), we find that
this transfer function can be directly derived, so long as the
scale factor as a function of time is given
\cite{grishchuk2000,zhang2005,kuro,tong2009,watanabe2006}. In this
paper, we shall use the following analytical approximation for
this transfer function. It has been known that, during the
expansion of the Universe, the mode function $h_k(\eta)$ of the
GWs behaves differently in two regions \cite{grishchuk2000}. When
waves are far outside the horizon, i.e. $k\ll aH$, the amplitude
of $h_k$ keeps constant, and when inside the horizon, i.e. $k\gg
aH$, the amplitude is damping with the expansion of Universe,
i.e., $h_k\propto 1/a(\eta)$. In the standard hot big-bang
cosmological model, we assume that the inflationary stage is
followed by a radiation dominant stage, and then the matter
dominant stage and the $\Lambda$ dominant stage. In this scenario,
by numerically integrating Eq.(\ref{h-evolution}), one finds that
the damping function $\dot{t}_f$ can be approximately described by
the following form
\cite{turner1993,zhao2006,efstathiou2006,giovannini2009}
 \begin{equation}\label{tf2}
 \dot{t}_f(\eta_0,k)=\frac{-3j_2(k\eta_0)\Omega_{m}}{k\eta_0}
 \sqrt{1+1.36(\frac{k}{k_{eq}})+2.50(\frac{k}{k_{eq}})^2},
 \end{equation}
where $k_{eq}=0.073\Omega_m h^2{\rm Mpc}^{-1}$ is the wavenumber
corresponding the Hubble radius at the time that matter and
radiation have equal energy density, and $\eta_0=1.41\times
10^{4}{\rm Mpc}$ is the present conformal time. The factor
$\Omega_m$ encodes the damping effect due to the recent
accelerating expansion of the Universe
\cite{zhang2005,kuro,zhao2006}. In this damping factor, we have
ignored the small effects of neutrino free-streaming
\cite{weinberg2003} and various phase transitions in the early
Universe \cite{watanabe2006}. In this paper, we shall focus on the
wavenumber $k\gg k_{eq}$. In this range, we have the current
density of RGWs as follows,
 \be\label{e71}
 \Omega_{\rm gw}(k) = \frac{15}{16} \frac{\Omega_m^2 A_s r}{H_0^2 \eta_0^4 k_{eq}^2}
\left(\frac{k}{k_*}\right)^{n_t},
 \ee
which clearly presents the dependence of the RGWs on various
cosmological parameters.

In the PTA analysis, people always describe the GW background by
the characteristic strain spectrum $h_c(f)$ \cite{maggiore}. For
most models of interest, it can be written as a power-law
dependence on frequency $f$:
 \be\label{hc}
 h_c(f)=A\left(\frac{f}{{\rm yr}^{-1}}\right)^{\alpha}.
 \ee
The characteristic strains relate to one-side power spectrum
$P(f)$ and the energy density of GWs $\Omega_{\rm gw}(f)$ as
 \be\label{pf}
 P(f)=\frac{h^2_c(f)}{12\pi^2f^3} ,~~\Omega_{\rm
 gw}(f)=\frac{2\pi^2}{3H_0^2}f^2h_c^2(f).
 \ee

Comparing the Equations in (\ref{e71}) and (\ref{pf}), we find
that
 \be\label{alpha}
 \alpha=\frac{n_t}{2}-1,
 \ee
and
 \be
 \frac{A}{\rm yr}=\sqrt{\frac{45}{32\pi^2}\frac{\Omega_m^2 A_s r}{\eta_0^4 k_{eq}^2}
\left(\frac{{\rm yr}^{-1}}{f_*}\right)^{n_t}}.
 \ee
Considering the cosmological parameters based on the current
Planck observations \cite{planck2013} $h=0.6711$,
$\Omega_m=0.3175$, $\Omega_{\Lambda}=0.6825$, $z_{eq}=3402$,
$A_s=2.495\times 10^{-9}$ at $k_*=0.002$Mpc$^{-1}$ \cite{ref32}, we obtain that
 \be\label{a}
 A=0.88\sqrt{r}\times 10^{5n_t-18},
 \ee
and
 \be\label{o}
 \Omega_{\rm gw}(f)=1.09r\times10^{10n_t-15}\left(f/{{\rm
 yr}^{-1}}\right)^{n_t}.
 \ee
These relations will be used for the following discussion. Both
Equations in (\ref{a}) and (\ref{o}) show that the amplitude of
RGWs at $f\sim 1/{\rm yr}$ strongly depends on the spectral index
$n_t$. For the cases with the scale-invariant and red spectrum,
one always has $A\lsim10^{-18}$ and $\Omega_{\rm
gw}\lsim10^{-15}$. However, for the cases with blue spectrum, i.e.
$n_t>0$, the values of $A$ and $\Omega_{\rm gw}$ can be
dramatically large. For example, in the models suggested by
Grishchuk \cite{grishchuk2005}, the blue spectrum with
$\alpha\in[-0.8,~-1]$ was expected, which corresponds to $n_t \in
[0,~0.4]$, the amplitude of RGWs could be $A\sim 10^{-17}$ and
$\Omega_{\rm gw} \sim 10^{-14}$.

\section{Pulsar Timing Array and the detection of relic gravitational waves\label{section3}}

\subsection{Current constraints}

\begin{figure}[t]
\centerline{\includegraphics[width=10cm,height=7cm]{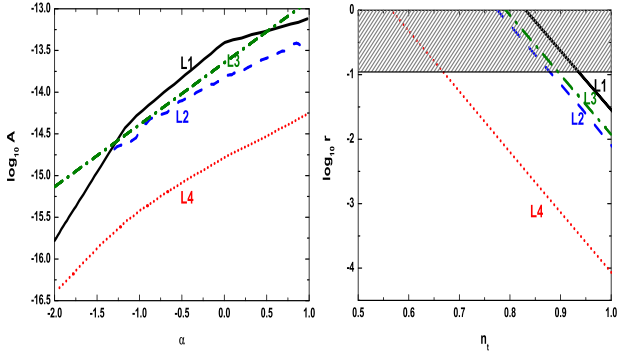}}
\caption{Left panel: The upper limit of $A$ as a function of the
spectral slope $\alpha$. Right panel: The upper limit of $r$ for
any given $n_t$, where the shaded region is excluded by the
current Planck observations. In each panel, the black solid line
(i.e. L1) is for current PPTA 2$\sigma$ result \cite{ppta}, the
blue dashed line (i.e. L2) is for current EPTA 2$\sigma$ result
\cite{epta}, the green dash-dotted line (i.e. L3) is for current
NANOGrav 2$\sigma$ result \cite{nano} and the red dotted line
(i.e. L4) is for future PPTA 2$\sigma$ result \cite{ppta}.
}\label{fig1}
\end{figure}

In 2006, Jenet et al. have analyzed the PPTA data and archival
Arecibo data for several millisecond pulsars. By focusing on the
GWs at the frequency $f=1/{\rm yr}$, the authors obtained the
2$\sigma$ upper limit on $A$ as a function of the spectral slope
$\alpha$, which is presented in the left panel of Fig.\ref{fig1}
(black solid line) \cite{ppta}. Recently, this upper limit has
been updated by EPTA and NANOGrav teams \cite{epta,nano}. It is
interesting that in \cite{ppta}, the authors have also
investigated the possible upper limit (or a definitive detection)
of stochastic background of GWs by using the potential completed
PPTA data-sets (20 pulsars with an rms timing residual of 100ns
over 5 years, which is also expected the case for future EPTA and
NANOGrav projects). We have also plotted the current EPTA upper
limit (blue dashed line), current NANOGrav upper limit (green
dash-dotted line) and the potential PPTA upper limit of parameter
$A$ in the left panel of Fig.\ref{fig1} (red dotted line).

By using the relations in Eqs. (\ref{alpha}) and (\ref{a}), we
obtain the constraints on the parameters $r$ and $n_t$, which are
presented in the right panel of Fig.\ref{fig1}. Note that the
regions above the lines are excluded by the corresponding PTA
observations. We notice that the current Planck observations give
the tightest constraint $r<0.11$ \cite{planck2013}, which is
nearly independent of the spectral index $n_t$ \cite{nt,nt2}. So,
combining with Planck constraint on $r$, this figure shows the
current allowed region in the $r$-$n_t$ plane. For example, if
$r=0.1$ is determined by the forthcoming CMB observations, current
PPTA gives the constraint $n_t<0.94$, NANOGrav gives $n_t<0.90$ 
and EPTA gives $n_t<0.88$ at
$2\sigma$ confident level. Meanwhile, the future PPTA will follow
the constraint of $n_t<0.67$. These are listed in Table
\ref{tab1}. Although quite loose, these constraints would be
helpful to exclude some inflationary models with very blue GW
spectrum.

From Fig.\ref{fig1} and Eq.(\ref{o}), we can also obtain the
constraints on the energy density of RGWs $\Omega_{\rm gw}(f)$. For
instance, if $n_t=0$ and $f=1/{\rm yr}$ are fixed, the upper limits
for $\Omega_{\rm gw}(f)$ are listed in Table \ref{tab2}, which are
consistent with the results in \cite{ppta,epta,nano}.

\begin{table*}
\caption{The 2$\sigma$ upper limit of the spectral index $n_t$
inferred from various pulsar timing observations. }
\begin{center}
\label{tab1}
\begin{tabular}{|c|c|c|c|c|c|c|c|}
    \hline
     & Current PPTA & Current EPTA & Current NANOGrav & Future PPTA & ~~~FAST ~~~& ~~~SKA~~~ & Optimal Case  \\
    \hline
   $r=0.1$ &   0.94   &  0.88    & 0.90 &    0.67        & 0.56& 0.32 & 0.07 \\
    \hline
   $r=0.01$ &   $>1$   &   0.99   & $>1$ &     0.78    & 0.67&  0.44 & 0.18 \\
    \hline
   $r=0.001$ &   $>1$  &  $>1$   & $>1$ &   0.89     &0.77 & 0.55 &  0.31 \\
   \hline
\end{tabular}
\end{center}
\end{table*}

\begin{table*}
\caption{The 2$\sigma$ upper limit of the energy density
$\log_{10}\left[\Omega_{\rm gw}(f={\rm yr}^{-1})\right]$ inferred
from various pulsar timing observations, where we have set
$n_t=0$, i.e. $\alpha=-1$. }
\begin{center}
\label{tab2}
\begin{tabular}{|c|c|c|c|c|c|c|}
    \hline
     Current PPTA & Current EPTA & Current NANOGrav & Future PPTA & ~~~FAST ~~~& ~~~SKA~~~ & Optimal Case  \\
    \hline
   -7.36   &  -7.79    & -7.63 &    -9.84        & -10.63& -12.99 &
   -15.30 \\
    \hline
\end{tabular}
\end{center}
\end{table*}

\subsection{Detecting GW background by Pulsar Timing Array}

In the following discussion, we shall study the potential
constraints on the RGWs by the future PTA observations, where we
will focus on the Chinese FAST project and the planned SKA
project.

The fluctuations of the pulsar TOAs caused by the stochastic GW
background are random. However, for different pulsars, these
fluctuations have the correlations. Let us assume the observations
of $n\gg 1$ pulsars at times $t_0$, $t_1$, ..., $t_{m-1}$ with the
time interval $\Delta t$. The total observation time is $T=m\Delta
t$. We denote the timing residual of $i$-th pulsar at time $t_k$
as $R^{i}_{k}$, which includes the contribution from both GWs
$s^{i}_{k}$ and the noises $n^{i}_{k}$, i.e.
$R_{k}^{i}=s_{k}^{i}+n_{k}^{i}$.

For the isotropic GW background, the correlation between the
GW-induced signals are \cite{hellings,jenet2005,lee2012}
 \be
 \langle s_{k}^{i} s_{k'}^{j}\rangle=\sigma_{g}^2
 {H_{ij}}\gamma_{kk'},
 \ee
where $\sigma_{g}$ is the root mean square (RMS) of the timing
residuals induced by GW background, which relates to the one-side
power spectrum $P(f)$ by $\sigma_{g}^2=\int_{f_{l}}^{f_{h}} P(f)
df$. The highest and lowest frequency of GWs are given by
$f_h=\frac{1}{2\Delta t}$ and $f_l=\frac{1}{T}$. $H_{ij}$ is the
so-called Hellings-Downs function, which is given by
$H_{ij}=\frac{3}{2}x\ln x-\frac{x}{4}+\frac{1}{2}(1+\delta(x))$,
where $x=\frac{1-\cos(\theta)}{2}$ and $\theta$ is the angle
distance between $i$-th and $j$-th pulsar. $\gamma_{kk'}$ is the
temporal correlation coefficient between the $k$-th and $k'$-th
sampling.

The noise term $n_{k}^{i}$ includes the effects of all non-GW
sources for the $i$-th pulsar. It is assumed that all noise
sources have a flat spectrum, which is consistent with most
observations \cite{jenet2005}. In order to simplify the problem,
in this paper, we assume all monitored pulsars have the same noise
level, i.e.
 \be
 \langle n_{k}^{i} n_{k'}^{j}\rangle=\sigma_{w}^2
 \delta_{ij}\delta_{kk'}.
 \ee

There are several methods to extract the GW signals from the
observable $R_k^i$ \cite{jenet2005,method1,method2}. In this
paper, we follow the method suggested by Jenet et al. in 2005
\cite{jenet2005}. In particular, we shall present the details of
the calculation, which are quite helpful to understand the method,
but have been neglected in the original paper \cite{jenet2005}. In
addition, some sub-dominant terms, which were neglected in
\cite{jenet2005}, will also be presented in the finial formulas.
We calculate the correlation coefficient between the observed
timing residuals of each pair of observed pulsars:
 \be
 c_{ij}=\frac{1}{m}\sum_{k=1}^{m} R_{k}^{i} R_{k}^{j}.
 \ee
It is easy to get the expected values of $c_{ij}$ and $c_{ij}^2$,
 \be\label{c}
 \langle c_{ij}\rangle = \sigma_g^2 H_{ij},
 \ee
 \be\label{c2}
 \langle c_{ij}^2\rangle = \sigma_g^4 \left(H^2_{ij}+\frac{(1+H_{ij}^2)\chi}{m}+\frac{2\sigma_w^2}{m\sigma_g^2}+\frac{4\sigma_w^4}{m\sigma_g^4}\right),
 \ee
where $\chi=\sum_{kk'}\gamma_{kk'}^2/m$, and $\langle\cdot\rangle$
denotes the ensemble average.

The comparison between $c_{ij}$ and the Hellings-Downs function is
carried out by defining the GW detection significance $S$ as
follows,
 \be\label{s}
 S=\frac{\sqrt{N}\sum_{i-j}(c_{ij}-\overline{c})(H_{ij}-\overline{H})}{\sqrt{\sum_{i-j}(c_{ij}-\overline{c})^2\sum_{i-j}(H_{ij}-\overline{H})^2}},
 \ee
where $N=n(n-1)/2$ is the number of independent pulsar pairs. The
summation $\sum_{i-j}$ sums over all independent pulsar pairs,
i.e. $\sum_{i-j}\equiv\sum_{i=1}^{n}\sum_{j=1}^{i-1}$. The
quantities $\overline{c}$ and $\overline{H}$ are defined as
 \be
 \overline{c}=\frac{1}{N}\sum_{i-j}c_{ij},~~\overline{H}=\frac{1}{N}\sum_{i-j}H_{ij}.
 \ee
To evaluate the quality of the detector, we need the expected
value $\langle S \rangle$, which is $\langle S\rangle
\simeq\sqrt{N}\sigma_g^2 \Sigma_{H}/\Sigma_{c}$, where
 \be
 \Sigma_{H}^2=\frac{1}{N}\sum_{i-j}(H_{ij}-\overline{H})^2,~~\Sigma_{c}^2=\frac{1}{N}\sum_{i-j}(c_{ij}-\overline{\langle
c\rangle})^2.
 \ee
By using Eqs. (\ref{c}) and (\ref{c2}), we get the well-known
result,
 \be\label{<s0>}
 \langle S\rangle
 \simeq\sqrt{N}\left[1+\frac{\chi(1+\overline{H^2})+\frac{2\sigma_w^2}{\sigma_g^2}+\frac{4\sigma_w^4}{\sigma_g^4}}{m\Sigma_H^2}\right]^{-1/2}.
 \ee

In Jenet et al. (2005), this formula was obtained by another way,
which is easier to extend to the results after low-pass filtering
and whitening. It is convenient to define the expected discrete
power spectrum of $R_{k}^{i}$ for the $i$-th pulsar
$P_d(\Delta,i)$, which includes both a GW component and a white
noise component, i.e. $P_d(\Delta,
i)=P_g(\Delta)+\frac{\sigma_w^2(i)}{m}$. Note that $\Delta>0$ is
the discrete frequency bin number corresponding to frequency
$\Delta/T$. Since we have assumed that $\sigma_w$ has the same
value for every pulsar, the spectrum $P_d(\Delta, i)$ becomes
independent of $i$, so we denote it as $P_d(\Delta)$ in the
following discussion. For the GW with the characteristic strain
spectrum $h_c(f)$ in Eq.(\ref{hc}), one has the discrete
GW-induced spectrum as follows,
 \be\label{pgdelta}
 P_g(\Delta)=\frac{(A\cdot{\rm yr})^2(T/{\rm
 yr})^{2-2\alpha}}{(2\pi)^2(2-2\alpha)} m(\Delta),
 \ee
where $m(\Delta=1)=\beta^{2\alpha-2}-1.5^{2\alpha-2}$, and
$m(\Delta>1)=(\Delta-0.5)^{2\alpha-2}-(\Delta+0.5)^{2\alpha-2}$.
$\beta\simeq 1$ is the lowest frequency used to calculate the
correlation function $c_{ij}$. According to the Wiener-Khinchin
theorem and the definition of $\Sigma_c$, we find that
 \beqa
 \Sigma_c^2  &=& \sigma_g^4\Sigma_H^2 + \sum_{\Delta} P^2_d(\Delta)+\overline{H^2}\sum_{\Delta}
 P_g^2(\Delta) \nonumber\\
 &=&\sigma_g^4\Sigma_H^2+(1+\overline{H^2})\sum_{\Delta}
 P_g^2(\Delta)+\sigma_g^4(\frac{2\sigma_w^2}{\sigma_g^2}+\frac{\sigma_w^4}{\sigma_g^4}),
 \nonumber
 \eeqa
and the quantity $\chi$ is calculated by
$\frac{\chi}{m}=\frac{1}{\sigma_g^4}\sum_{\Delta} P_g^2(\Delta)$,
which can be gotten for any given GW background. By using the
relation $\langle S\rangle \simeq\sqrt{N}\sigma_g^2
\Sigma_{H}/\Sigma_{c}$, we can naturally obtain the result in Eq.
(\ref{<s0>}).

In order to enhance the detection significance, the low-pass
filtering and whitening techniques can be applied
\cite{jenet2005}. In this way, we can correlate only that part of
signal which has a high signal-to-noise ratio and give each time
series a flat spectrum to optimize the measurement of the
correlation function. In practice, we define the new discrete
power spectrum $\hat{P}_d(\Delta)$ and $\hat{P}_g(\Delta)$ as
follows,
 \be
 \hat{P}_d(\Delta)=\frac{P_d(\Delta)}{P_d(\Delta)}\frac{\sigma_d^2}{m},~~
 \hat{P}_g(\Delta)=\frac{P_g(\Delta)}{P_d(\Delta)}\frac{\sigma_d^2}{m},
 \ee
where $\sigma_d^2=\sum_{\Delta} P_d(\Delta)$. In this definition,
the total RMS fluctuation induced by GW becomes
$\hat{\sigma}_g^2=\sum_{\Delta=1}^{\Delta_{\max}}\hat{P}_g(\Delta)$,
where the summation is carried out only over the frequency bins in
which the GW signal dominates the noise, and $\Delta_{\max}$ is
the number of the highest frequency bin. So the variance
$\Sigma_c$ becomes
 \beqa
 \Sigma_c^2&=&\hat{\sigma}_g^4\Sigma_H^2+\sum_{\Delta} \hat{P}^2_d(\Delta)+\overline{H^2}\sum_{\Delta}
 \hat{P}_g^2(\Delta) \\
 &=&\hat{\sigma}_g^4\Sigma_H^2+\frac{\sigma_d^4}{m^2}\left[\sum_{\Delta=1}^{\Delta_{\max}}\left(1+(\frac{P_g(\Delta)}{P_d(\Delta)})^2\overline{H^2}\right)\right],
 \eeqa
and the expected value of $S$ becomes
 \be\label{<s>}
 \langle S\rangle
 \simeq\sqrt{N}\left[1+\frac{\sum_{\Delta=1}^{\Delta_{\max}}\left(1+(\frac{P_g(\Delta)}{P_d(\Delta)})^2\overline{H^2}\right)}
 {(\sum_{\Delta=1}^{\Delta_{\max}}\frac{P_g(\Delta)}{P_d(\Delta)})^2 \Sigma_H^2}\right]^{-1/2}.
 \ee
This formula will be used in the following subsection.

\subsection{Forecasts for FAST and SKA projects}

FAST is a Chinese megascience project to build the largest single
dish radio telescope in the world. Funding for FAST has been
approved in 2007, and its first light is expected to be in 2016
\cite{fast}. It includes multibeam and multiband, covering a
frequency range of 70MHz$-$3GHz. The relatively low latitude
($\sim 26^{\circ}$N) of the site enables the observation of more
southern galactic pulsars. The zenith angle of FAST is about
$40^{\circ}$, which corresponds to $\overline{H^2}=0.024$ and
$\Sigma_{H}=0.155$, if assuming the monitored millisecond pulsars
evenly distribute in the observed region. One of the scientific
goals of FAST is to discover $\sim 400$ new millisecond pulsars.
FAST is capable of providing the most precise observations of
pulsar timing signals, therefore, may largely increase the
sensitivity of the spectrum window for detection of GWs.

\begin{figure}[t]
\centerline{\includegraphics[width=10cm,height=7cm]{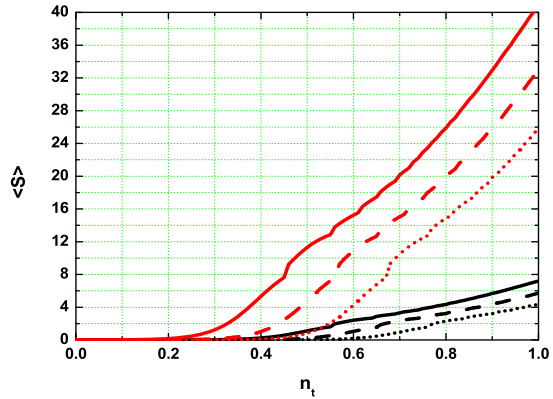}}
\caption{The detection significance of RGWs for FAST (dark lines, black online)
and SKA (gery lines, red online) projects. For FAST, we have assumed $T=5$yr,
$\sigma_w=30$ns, and $n=20$, and for SKA we have assumed $T=10$yr,
$\sigma_w=50$ns, and $n=100$. For both cases, solid lines are for
the models with $r=0.1$, dashed lines are for $r=0.01$, and dotted
lines are for $r=0.001$.}\label{fig2}
\end{figure}

\begin{figure}[t]
\centerline{\includegraphics[width=10cm,height=7cm]{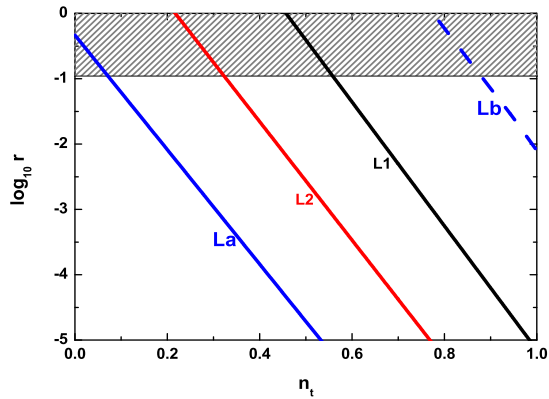}}
\caption{The upper limits of $r$ and $n_t$ based on the potential
FAST observations (black line, i.e. L1) and SKA observations (red
line, i.e. L2). Note that the shaded region is excluded by current
Planck observations. The dashed blue line (i.e. Lb) is the current
tightest constraint coming from the EPTA observations, which is
identical to that in the right panel of Fig. \ref{fig1}. The solid
blue line (i.e. La) indicates the result in the optimal case
considered in this paper, where $T=20$yr, $\sigma_w=30$ns, and
$n=200$ are assumed.}\label{fig3}
\end{figure}

The noise level of the millisecond pulsars are expected to be
$\sigma_w=30$ns, after collecting the timing data for the total
time $T=5$yr \cite{fast}. As a conservative evaluate, similar to
PPTA, we assume FAST will monitor 20 pulsars for the detection of
GWs. Thus, by using Eq.(\ref{<s>}), we can calculate the
detection significance $S$ for any given RGW models, which are
illustrated in Fig.\ref{fig2}. In this figure, we have considered
three typical models with $r=0.1$, $0.01$ and $0.001$. These
models are predicted by the general inflationary models, and could
be well detected by the future CMB observations \cite{zhao}. As
anticipated, if $n_t<0$, we always have $S \ll 1$, i.e. the
detection is impossible for the red spectrum of RGWs. However, if
the RGWs have the blue spectrum, the detection is possible. For
instance, for the model with $r=0.1$ and $n_t=0.56$ or for that
with $r=0.01$ and $n_t=0.67$, FAST can detect the signal of RGWs
at 2$\sigma$ level. In Fig.\ref{fig3}, we set $\langle
S\rangle=2$, and plot the value of $r$ for any spectral index
$n_t$. Comparing those in the right panel of Fig.\ref{fig1}, we
find that FAST is much more sensitive than current and future PPTA
and/or EPTA.

As another potential observation, we consider SKA project, which
is a proposed major internationally-funded radio telescope, and is
expected to be completed in the next decade \cite{ska1}. SKA will
consist of many antennas, constituting an effective collecting
area of about one square kilometer. We expect that SKA will survey
the full sky. If assuming the monitored millisecond pulsars are
evenly distributed, we have $\overline{H^2}=\Sigma^2_{H}=1/48$,
which are slightly different from those of FAST. Following
\cite{ska2}, we assume SKA will select $100$ pulsars and spend the
total time $T=10$yr for the GW detection, and the average noise
level of these pulsars are about $\sigma_w=50$ns, which is 2 times
lower than those the finial PPTA, EPTA or NANOGrav. In
Fig.\ref{fig2}, we consider the typical inflationary models with
$r=0.1$, $0.01$ and $0.001$, and plot the values of $S$ for any
$n_t$. Again, we find the detection is possible, only if $n_t>0$,
i.e. the blue GW spectrum. Compared with the results of FAST, the
detection significance are much higher, due to the longer
observation time $T$ and the larger pulsar number $n$. These are
also clearly shown in Fig.\ref{fig3} and Table \ref{tab1}. In Table
\ref{tab2}, we have listed the detection limits of the energy density 
$\Omega_{\rm gw}$ for the FAST and SKA projects, where we also find
that SKA is more sensitive than FAST.

\begin{figure}[t]
\centerline{\includegraphics[width=10cm,height=7cm]{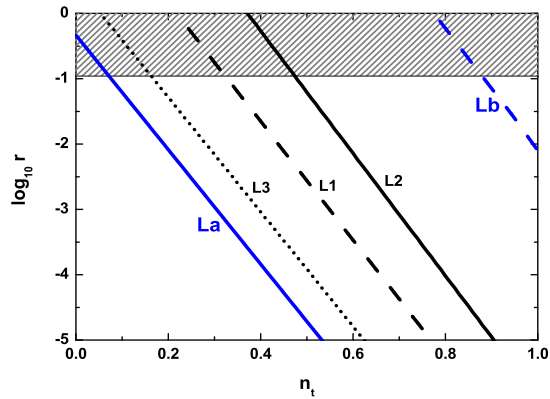}}
\caption{The upper limits of $r$ and $n_t$ depend on the total
observation time $T$. The solid black line (i.e. L2) is for the
case with $T=5$yr, dashed black line (i.e. L1) is for $T=10$yr,
and dotted black line (i.e. L3) is for $T=20$yr. In all cases,
$\sigma_w=50$ns and $n=100$ are assumed.The solid blue line (i.e.
La) and dashed blue line (i.e. Lb) are identical to those in
Fig.\ref{fig3}. The dashed black line (i.e. L1) is identical to that for
SKA.}\label{fig4}
\end{figure}

From the formula in Eq.(\ref{<s>}), we know that the detection
significance of PTA projects mainly depends on three factors: the
total observation time $T$, the number of the monitored
millisecond pulsars $n$ and the noise level of the pulsar
$\sigma_w$ \cite{foot1}. Now, let us discuss the dependence of
sensitivity on these factors separately. First, we fix
$\sigma_w=50$ns and $n=100$, and investigate the effect of
observation time $T$. To do it, we consider three cases with
$T=5$yr, $10$yr and $20$yr. Setting the detection significance
$\langle S\rangle=2$, we plot the constraints of the inflationary
models in the $r$-$n_t$ plane in Fig.\ref{fig4}, where we find the
effect of total time $T$ is very significant. For example, for the
model with $r=0.1$, the 5yr observations give the constraint
$n_t<0.48$, which can be improved to $n_t<0.17$ for the 20yr
observations. For comparison, in this figure, we have also
consider the optimal case, where $\sigma_w=30$ns, $n=200$ and
$T=20$yr are assumed. We find that, in this optimal case, the
constraint of spectral index is only slightly improved to
$n_t<0.07$, although the noise level and pulsar numbers are
greatly improved.

This effect can be understood by the following analysis. As well
known, the contributions of GWs on the pulsar timing residuals
mainly come from those at the lowest frequency range, i.e. $f\sim
f_l$. So the detection significance $S$ sensitively depends on the
$f_l$ value. At the same time, we know that $f_l=1/T$. So the
larger total observation time $T$ corresponds to the smaller $f_l$
value, which means that more low-frequency GWs can contribute the
timing residuals of pulsars. This explains why the observation
time $T$ is the most important factor for the sensitivity of PTA.

\begin{figure}[t]
\centerline{\includegraphics[width=10cm,height=7cm]{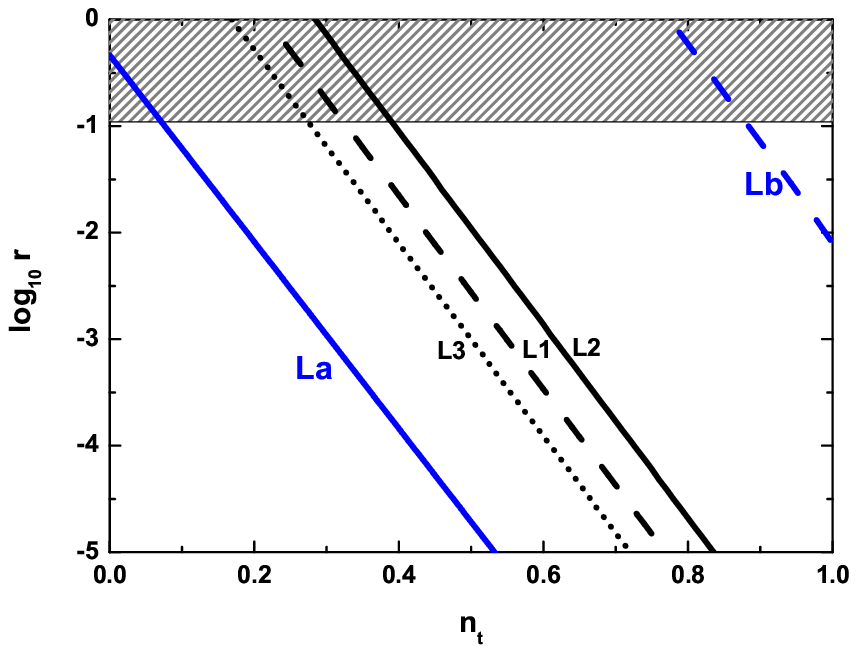}}
\caption{The upper limits of $r$ and $n_t$ depend on the noise
level $\sigma_w$. The solid black line (i.e. L2) is for the case
with $\sigma_w=100$ns, dashed black line (i.e. L1) is for
$\sigma_w=50$ns, and dotted black line (i.e. L3) is for
$\sigma_w=30$ns. In all cases, $T=10$yr and $n=100$ are assumed.
The solid blue line (i.e. La) and dashed blue line (i.e. Lb) are
identical to those in Fig.\ref{fig3}. The dashed black line (i.e. L1) is
identical to that for SKA.}\label{fig5}
\end{figure}

Second, we study the effect of noise level of pulsars $\sigma_w$.
Decreasing $\sigma_w$ is equivalent to increasing the
$\Delta_{\max}$ value. So a smaller $\sigma_w$ corresponds to the
case where more high-frequency GWs have the contributions to the
pulsar timing residuals. However, we know that, compared with the
low-frequency GWs, the high-frequency ones are much less important
for the timing residuals. The results are shown in Fig.\ref{fig5},
where three cases with $\sigma_w=100$ns, $50$ns and $30$ns are
considered. Although as anticipated, lower $\sigma_w$ corresponds
to the higher sensitivity of PTA, the effect of $\sigma_w$ is less
significant than that of observation time $T$.

\begin{figure}[t]
\centerline{\includegraphics[width=10cm,height=7cm]{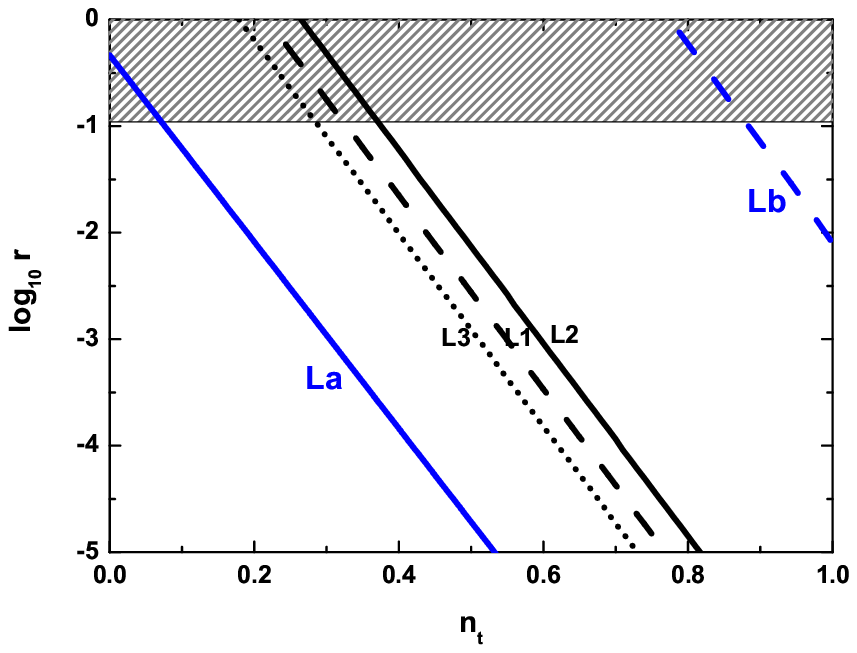}}
\caption{The upper limits of $r$ and $n_t$ depend on the monitored
pulsar number $n$. The solid black line (i.e. L2) is for the case
with $n=50$, dashed black line is for $n=100$ (i.e. L1), and
dotted black line is for $n=200$ (i.e. L3). In all cases,
$\sigma_w=50$ns and $T=10$yr are assumed. The solid blue line
(i.e. La) and dashed blue line (i.e. Lb) are identical to those in
Fig.\ref{fig3}. The dashed black line (i.e. L1) is identical to that for
SKA.}\label{fig6}
\end{figure}

Third, the pulsar number $n$ affects the value of $S$ only by the
factor $\sqrt{N}$ in Eq.(\ref{<s>}), which follows that $\langle
S\rangle \propto n$ for $n\gg 1$. This effect is illustrated in
Fig.\ref{fig6}, where three cases with $n=50$, $100$ and $200$ are
considered. We find that, compared with the total observation time
$T$ and the pulsar noise level $\sigma_w$, the pulsar number $n$
has the relatively smaller influence on the detection significance
$S$.

\section{Conclusions\label{section4}}

Generation of GW background in the early inflationary stage is a
necessity dictated by general relativity and quantum mechanics.
The wide range spreading spectra of RGWs make the possible
detection at different frequency ranges by various methods. The
timing studies of the millisecond pulsars provide a unique way to
constrain it in the middle frequency range
$f\in(10^{-9},~10^{-7})$Hz.

Recently, PPTA, EPTA and NANOGrav teams have reported their observational
results on GW background at $f\sim 1/{\rm yr}$. In this paper, we
infer from these bounds the constraint of inflation in $r$-$n_t$
plane. Although quite loose, these constraints are helpful to
exclude some phantom-like inflationary models.

As the main goal of this paper, we have forecasted the future
pulsar timing observations and the potential constraints on
inflationary parameters $r$ and $n_t$, by focusing on the FAST and
SKA projects. We found that, if $r=0.1$, FAST could give the
constraint on the spectral index $n_t<0.56$, and SKA gives
$n_t<0.32$. While an observation with the total time $T=20$yr, the
pulsar noise level $\sigma_w=30$ns and the monitored pulsar number
$n=200$, could even constrain $n_t<0.07$, which can exclude or
test most phantom-like inflationary models with this
tensor-to-scalar ratio. In this paper, we have also studied the
effects of $T$, $\sigma_w$ and $n$ on the sensitivity of PTA, and
found that the total observation time $T$ has the most important
influence. So increasing the observation time can significantly
improve the sensitivities of the future PTAs.

~


{\bf Acknowledgements:} We appreciate helpful discussion with K.J. Lee and D. Li. This work is supported by the Ministry of
Science and Technology National Basic Science Program (Project
973) under Grant No.2012CB821804. WZ is supported by NSFC No.
11173021, 11075141 and project of Knowledge Innovation Program of
CAS. YZ is supported by NSFC No. 10773009, SRFDP, and CAS. XPY is
supported by NSFC No. 10803004, CQ CSTC No. 2008BB0265 and the
Fundamental Research Funds for the Central Universities
(XDJK2012C043). ZHZ is supported by NSFC No.11073005, 
the Fundamental Research Funds for the Central Universities and Scientific Research Foundation of Beijing
Normal University.


\end{document}